\documentclass{article}
\usepackage{spconf,amsmath,graphicx}
\usepackage{enumitem,caption,float}
\setlist{nosep, leftmargin=14pt}
\usepackage{mwe} 
\usepackage{cuted}
\usepackage{color}

\title{ACCELERATION: Sequenti\underline{a}lly-S\underline{c}anning Dual-Energy \underline{C}omputed Tomography Imaging Using High T\underline{e}mpora\underline{l} R\underline{e}solution Image Reconst\underline{r}uction and Tempor\underline{a}l Extrapola\underline{tion}}
\name{Qiaoxin Li, Dong Liang, Yinsheng Li}
\address{Research Center for Medical Artificial Intelligence, \\ 
	Shenzhen Institutes of Advanced Technology, Chinese Academy of Sciences, Shenzhen, China \\
	University of Chinese Academy of Sciences, Beijing, China}

\begin{document}
	
	\maketitle
	
	\begin{abstract}
		
		Dual-energy computed tomography (DECT) has been widely used to obtain quantitative elemental composition of imaged subjects for personalized and precise medical diagnosis. Compared with existing high-end DECT leveraging advanced X-ray source and/or detector technologies, the use of the sequentially-scanning data acquisition scheme to implement DECT may make broader impact on clinical practice because this scheme requires no specialized hardware designs. However, since the concentration of iodinated contrast agent in the imaged subject varies over time, sequentially-scanned data sets acquired at two tube potentials are temporally inconsistent. As existing material decomposition approaches for DECT assume that the data sets acquired at two tube potentials are temporally consistent, the violation of this assumption results in inaccurate quantification accuracy of iodine concentration. In this work, we developed a technique to achieve sequenti\underline{a}lly-s\underline{c}anning DE\underline{C}T imaging using high t\underline{e}mpora\underline{l} r\underline{e}solution image reconst\underline{r}uction and tempor\underline{a}l extrapola\underline{tion}, ACCELERATION in short, to address the technical challenge induced by temporal inconsistency of sequentially-scanned data sets and improve iodine quantification accuracy in sequentially-scanning DECT. ACCELERATION has been validated and evaluated using numerical simulation data sets generated from clinical human subject exams. Results demonstrated the improvement of iodine quantification accuracy using ACCELERATION. 
		%XXX first reconstructs several time-resolved images using the short-scan data acquired at the first tube potential. The imaged subject corresponding to the acquisition time of the second tube potential can be obtained via temporally extrapolating the time-resolved images corresponding the first tube potential. Iodine concentration can be quantified using the pair of calculated temporally consistent dual-energy images.
	\end{abstract}
	
	\begin{keywords}
		Dual-Energy Computed Tomography; Sequential Scan Scheme; Temporal Inconsistency; Material Decomposition; Deep Learning
	\end{keywords}
	
	\section{Introduction}
	
	%Dual-energy computed tomography (DECT) has been developed to address one of the major technical challenges of conventional CT: The linear attenuation coefficient obtained by conventional CT depends on both intrinsic characteristic quantities of an imaged subject (i.e., elemental composition) and extrinsic data acquisition conditions (i.e., energy spectrum of a given data acquisition scheme of a given CT). As a result, conventional CT fails to quantify the elemental composition of a subject for personalized and precise medical diagnosis.

	\begin{figure}[ht]
	\centering
	\includegraphics[width=0.4\textwidth, keepaspectratio=true]{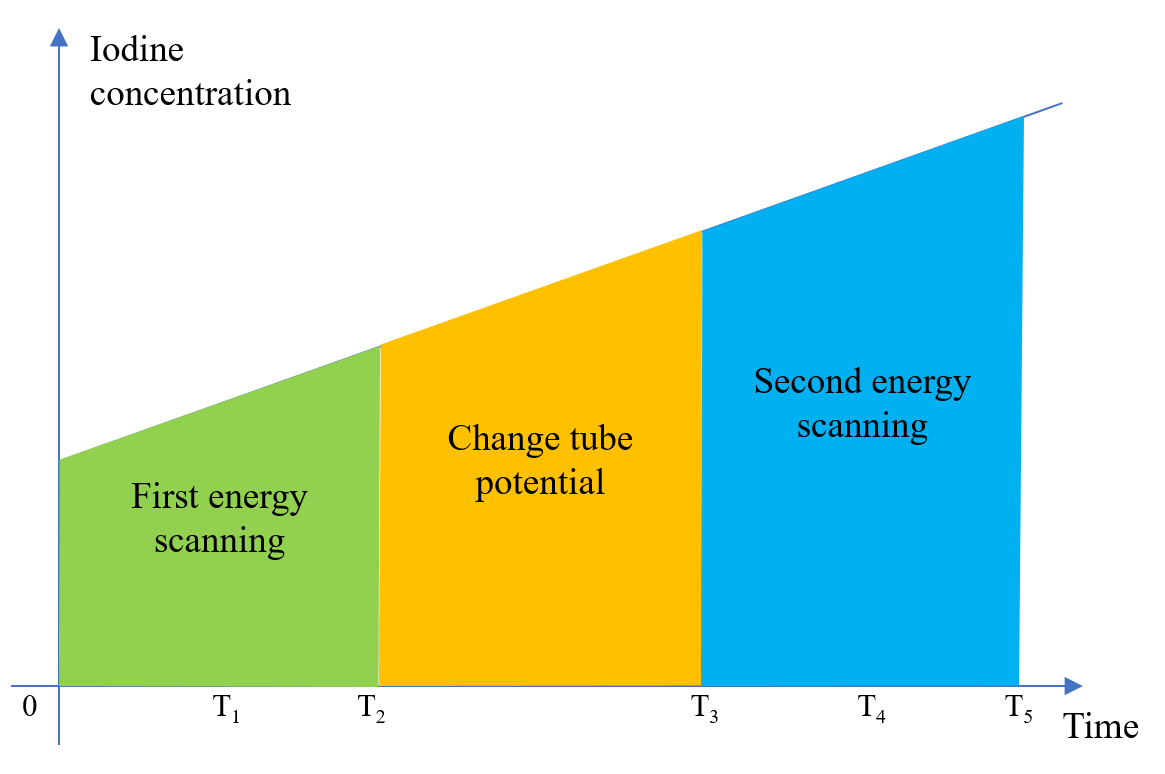}
	\caption{Temporal inconsistency in the sequentially-scanning DECT data acquisition.}
	\label{figure1}
	\end{figure}

	%To address the technical challenge, 
	Dual-energy computed tomography (DECT) has been widely used to obtain quantitative elemental composition of imaged subjects for personalized and precise medical diagnosis. Significant efforts have been made in clinical DECT to acquire temporally consistent data sets at two different tube potentials leveraging advanced X-ray source and/or detector technologies \cite{mccollough2020principles}. As these high-end DECT systems may be limited to major clinical centers or research institutes, the use of the sequentially-scanning data acquisition scheme to implement DECT may make broader impact on clinical practice because this scheme requires no specialized hardware designs. In sequentially-scanning DECT, data sets corresponding to two different tube potentials are acquired at two different time points. Since the concentration of iodinated contrast agent in the imaged subject varies over time, as demonstrated by Fig.~\ref{figure1}, sequentially-scanned data sets acquired at two tube potentials are temporally inconsistent. As existing material decomposition approaches for DECT assume that the data sets acquired at two tube potentials are temporally consistent, the violation of this assumption results in inaccurate quantification accuracy of iodine concentration. 
	
	To address the technical challenge induced by temporal inconsistency of sequentially-scanned data sets and improve iodine quantification accuracy in sequentially-scanning DECT, in this work, we developed a technique to achieve sequenti\underline{a}lly-s\underline{c}anning DE\underline{C}T imaging using high t\underline{e}mpora\underline{l} r\underline{e}solution image reconst\underline{r}uction and tempor\underline{a}l extrapola\underline{tion}, ACCELERATION in short. ACCELERATION first reconstructs several time-resolved images, each corresponding to an improved temporal resolution, using the short-scan data acquired at the first tube potential. The imaged subject corresponding to the acquisition time of the second tube potential can be obtained via temporally extrapolating the time-resolved images corresponding the first tube potential. Iodine concentration can be quantified using the pair of calculated temporally consistent images at two tube potentials. ACCELERATION has been validated and evaluated using numerical simulation data sets generated from clinical human subject exams. Results demonstrated the improvement of iodine quantification accuracy using ACCELERATION. 

	\section{The Proposed ACCELERATION Technique}
	
	Our goal is to address the technical challenge induced by temporal inconsistency of sequentially-scanned data sets and improve iodine quantification accuracy in sequentially-scanning DECT. As shown in Fig.~\ref{figure1}, suppose that scanning at the first tube potential is performed during $[0,T_{2}]$, the tube potential is changed during $[T_{2},T_{3}]$, and scanning at the second tube potential is performed during $[T_{3},T_{5}]$. The variation of iodine concentration in imaged subjects underwent intravenous contrast injection at each image voxel can be approximately considered as general Gamma-Variate functions \cite{preslar1991new}. \textbf{Given the short duration of data acquisition (e.g. about 0.2 seconds to acquire a short-scan data set using a state-of-the-art CT scanner), the variation of iodine concentration within this short duration can be legitimately assumed as linear.}
	
	The input of ACCELERATION is the short-scan sinograms at different tube potentials acquired during $[0,T_{2}]$ and $[T_{3},T_{5}]$, and its output is the reconstructed CT images at the same time (referred to as $T_{4}$ in this paper) at two tube potentials and its decomposed water basis and iodine basis images. 
	
	\subsection{Reconstruction by solving Eq. (\ref{eq1})}
	
\begin{figure}[H]
		\centering
		\includegraphics[scale=0.4]{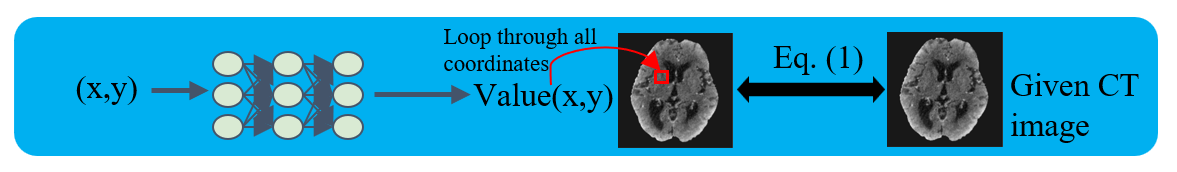}
		\caption{Reconstruction by solving Eq. (\ref{eq1})}
		\label{figure2}
	\end{figure}

	In implicit neural representation learning \cite{molaei2023implicit}, a multi-layer perceptron (MLP) $\mathcal{M}_{\theta}$ is trained to express the mapping relationship between pixel coordinates and pixel values. That is, $\mathcal{M}_{\theta}(x_i) \rightarrow v_i$, where $i$ denotes the coordinate index in the image spatial space. The optimization objective can be formulated as follows:
	\begin{equation}
		{\theta}^{*}=argmin_{\theta} \frac{1}{N}\sum_{j=1}^{N} \left(\mathcal{M}_{\theta}(x_j) - v_j\right) ^{2},
	\label{eq1}
	\end{equation}
	where $N$ denotes the total number of voxels in the CT image. After training, looping over all coordinates can obtain the input CT image (see in Fig.~\ref{figure2}).
	
	\subsection{Reconstruction by solving Eq. (\ref{eq2})}
	
\begin{figure}[H]
		\centering
		\includegraphics[scale=0.4]{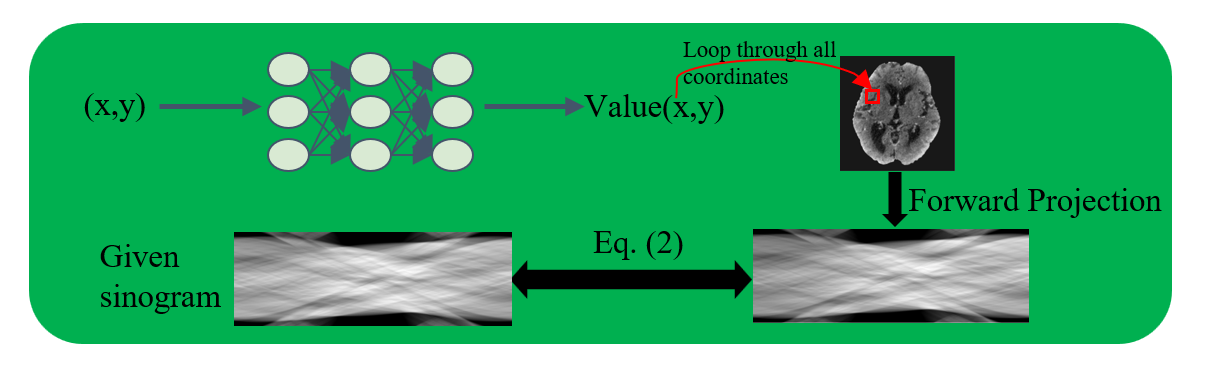}
		\caption{Reconstruction by solving Eq. (\ref{eq2})}
		\label{figure3}
	\end{figure}

	In this case, we use sinogram as the label (see in Fig.~\ref{figure3}). That is, we design the loss function in the sinogram domain \cite{shen2022nerp} after forward projecting the output image of MLP. The optimization objective is as follows:
	\begin{equation}
		{\theta}^{*}=argmin_{\theta} \frac{1}{M}\sum_{i=1}^{M} \left( \sum_{j=1}^N a_{ij} \mathcal{M}_{\theta}(x_j) - s_i \right)^2,
	\label{eq2}
	\end{equation}
	where $M$ represents the number of measurements in the sinogram space, $a_{ij}$ represents $i$-th, $j$-th element in the forward projection operator and $s_i$ represents the $i$-th measured datum in the sinogram, respectively.
	
	\subsection{Short-scan image reconstruction}
	
\begin{figure}[H]
		\centering
		\includegraphics[scale=0.31]{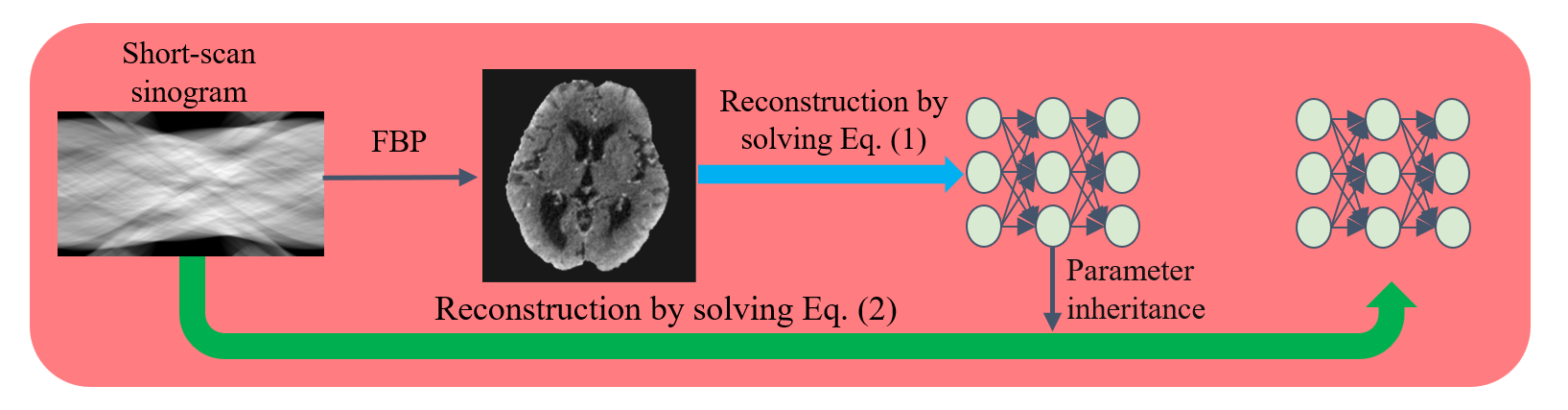}
		\caption{Short-scan image reconstruction module}
		\label{figure4}
	\end{figure}

	In this section, we use Eq. (\ref{eq1}) and Eq. (\ref{eq2}) to train in sequence in the image domain and sinogram domain to obtain accurate CT reconstruction images (see in Fig.~\ref{figure4}). The image obtained by FBP is used to obtain the MLP parameter $\theta_1$ using Eq. (\ref{eq1}). Then, the sinogram and $\theta_1$ are used to obtain $\theta_2$ using Eq. (\ref{eq2}) (which can be regarded as fine-tuning of $\theta_1$) .
	
	\subsection{ACCELERATION}
	\begin{figure*}
		\centering
		\includegraphics[width=0.95\textwidth, keepaspectratio=true]{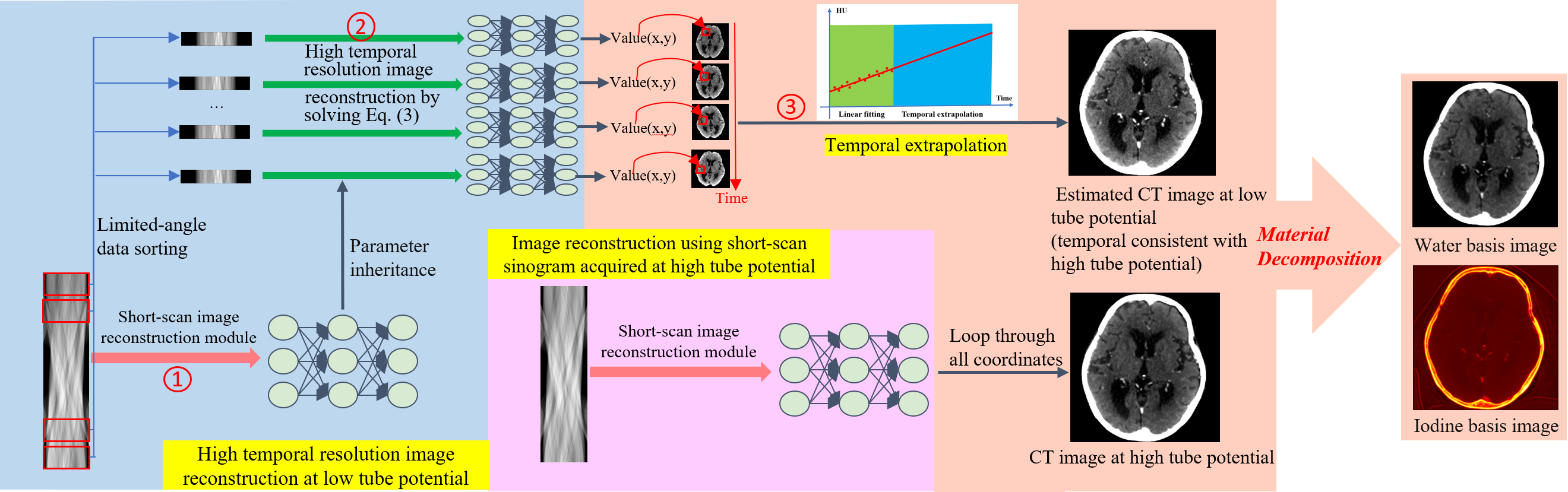}
		\caption{Global architecture of ACCELERATION. \normalsize{\textcircled{\scriptsize{1}}}\normalsize, \normalsize{\textcircled{\scriptsize{2}}}\normalsize, \normalsize{\textcircled{\scriptsize{3}}}\normalsize $~$represent the first, second and third steps respectively for the data set acquired at the low tube potential.}
		\label{figure5}
	\end{figure*}
	
	Based on the three parts introduced above, we developed ACCELERATION technique to address the technical challenge induced by temporal inconsistency of sequentially-scanned data sets and improve iodine quantification accuracy in sequentially-scanning DECT.
	
	At the first tube potential (see in pale blue section in Fig.~\ref{figure5}), in the first step, short-scan data is input into Short-scan image reconstruction module (see in \ref{figure4}) for reconstruction, obtaining the image and its corresponding MLP parameter $\theta_0$ at time $T_{1}$ (see in Fig.~\ref{figure1}). In the second step, sorting limited-angle data. The optimization objective can be written as follows:
	\begin{align}
		{\theta_t}^{*}&=argmin_{\theta_t} \frac{1}{|S_t|}\sum_{i\in S_t} \left( \sum_{j=1}^N a_{ij} \mathcal{M}_{\theta_t}(x_j) - s_i \right)^2, \\ \nonumber
		S_t &= \left\lbrace \frac{M}{T}(t-1)+1, \frac{M}{T}(t-1)+2, \cdots, \frac{M}{T}t \right\rbrace, \\ \nonumber
	\end{align}
	where $\theta_t$ denotes the MLP parameters corresponding to the $t$-th time frame, $t = 1,2,\cdots,T$. $S_t$ denotes the set of measurement corresponding to the $t$-th time frame, $|\cdot|$ denotes the number of elements in the set, and $T$ denotes the number of reconstructed time frames which is a hyper-parameter to be optimized for the best reconstruction accuracy.
	
	According to the physical process of CT data acquisition, the $T$ segments have a linear temporal sequence, thus the CT images obtained by different segments also have the temporal sequence (uniformly distributed in $[0,T_{2}]$). In the third step, based on the $T$ CT images obtained in the previous step, for each pixel in the CT, a linear least squares fitting is used, and then extrapolated to time $T_4$ to predict the CT image at time $T_4$ at first tube potential (see in pale orange section in Fig.~\ref{figure5}). The scan data at the second tube potential only needs to perform the same first step as the first tube potential (see in lavender section in Fig.~\ref{figure5}) at $T_4$. 
	
	After the above operations, the temporal inconsistency in sequentially-scanning DECT is mitigated, and we can obtain CT images at time $T_4$ at different tube potentials, and then perform material decomposition to obtain accurate material basis images.

	\section{Materials}
	
	Material basis images (water-iodine basis pair) of a human subject who underwent the dual-energy contrast-enhanced cerebral CT angiographic imaging were used in the simulation. The iodine concentration increases by 0.5 times of the original value within a short scan(see $[0, T_2]$, $[T_3, T_5]$ in  Fig.~\ref{figure1}). The time taken to change the tube potential is the same as the time taken for a short scan. Therefore, the iodine concentration changes to 2.5 times the collected data in the interval $[0, T_5]$. Assuming a short-scan contains 1000 views, the iodine concentration at each view time within $[0, T_5]$ can be interpolated based on the linear variation of the iodine concentration. The CT images at different tube potentials at different times can be calculated based on the water concentration and iodine concentration at corresponding times. By performing a forward projection at corresponding view on each simulated CT image, we can simulate the sinograms of the two short scans $[0, T_2]$ and $[T_3, T_3]$. Based on these sinograms, we can proceed with the reconstruction and material decomposition, and the simulated CT images can then be used for validation.
	
	\section{Results}
	
	\subsection{Algorithm Performance Analysis}
	\begin{figure}[t]
		\centering
		\includegraphics[scale=0.6]{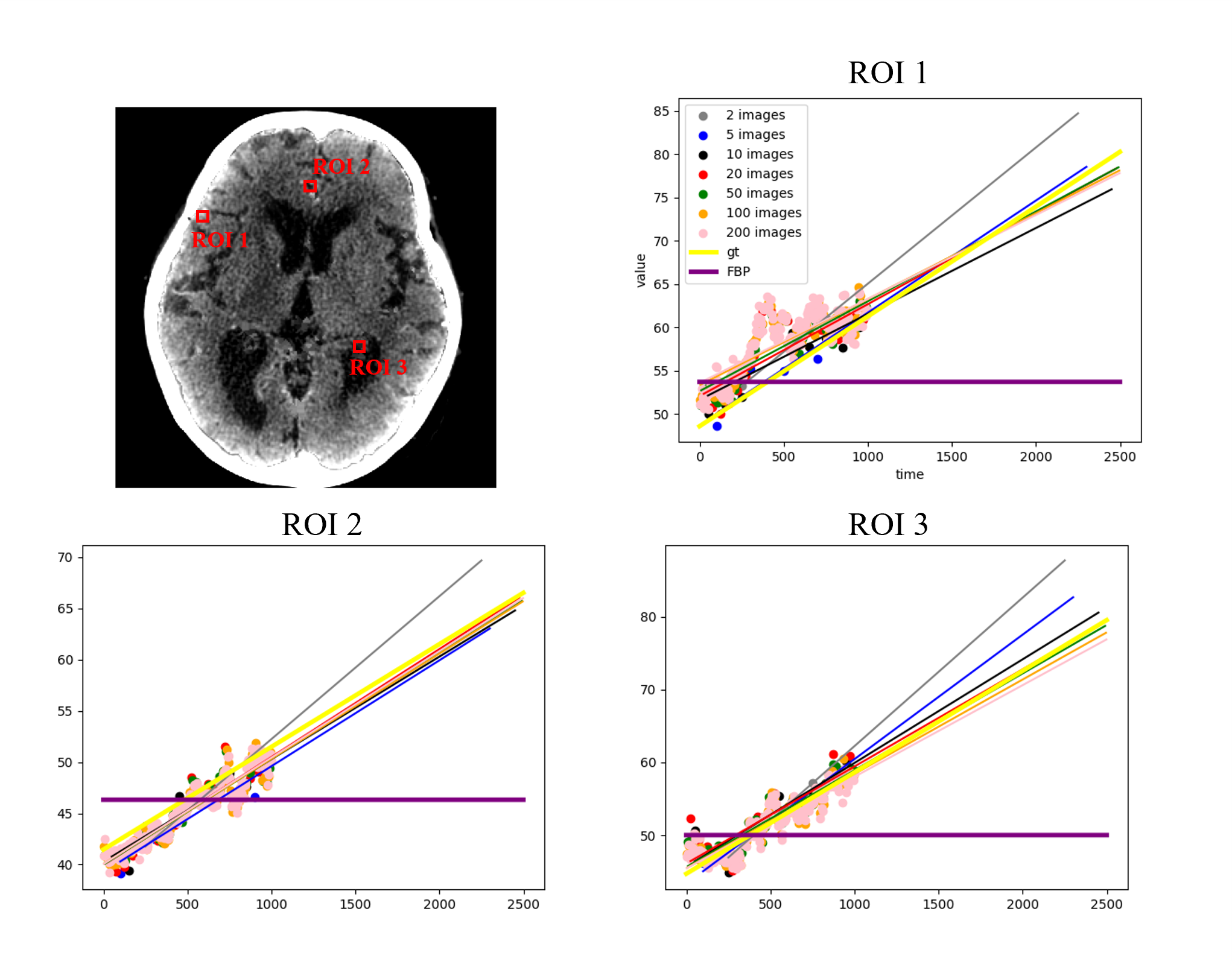}
		\caption{Scatter plot and fitted line graph of the reconstructed images in different ROIs, where the purple line represents the CT values obtained from FBP without extrapolation prediction, the yellow line represents the ground truth, and the other colors represent the CT value changes fitted by different numbers of reconstructed images.Window size of ROI is $3\times3$ pixels.}
		\label{figure6}
	\end{figure}
	The performance of ACCELERATION is largely influenced by the number of reconstructed images(i.e., the $T$ in section 2.4). This section mainly analyzes the impact of the number of reconstructed images on the accuracy of the predicted image (the CT image at first tube potential extrapolated to time $T_{4}$).Table.~\ref{table1} shows that the image accuracy increases rapidly and then decreases slowly as the number of reconstructed images increases. 
	
	\begin{table}[h]
		\centering
		\begin{tabular}{ccccc}
			\hline
			Image Number & 0(FBP) & 2 & 5 & 10 \\
			\hline
			NRMSE & 0.4457 & 0.0240 & 0.0169 & 0.0153 \\
			\hline
			Image Number & 20 & 50 & 100 & 200 \\
			\hline
			NRMSE & 0.0148 & 0.0148 & 0.0150 & 0.0151 \\
			\hline
		\end{tabular}
		\caption{NRMSE different reconstructed image number}
		\label{table1}
	\end{table}

	From the perspective of pixel-level accuracy, Fig.~\ref{figure6} shows the linear changes in CT fitted by different numbers of reconstructed images in different ROIs. It can be seen from the figure that ACCELERATION can correctly predict the change trend of each ROI and has good accuracy.

	\subsection{Material Decomposition}
	\begin{figure}[t]
		\centering
		\includegraphics[scale=0.55]{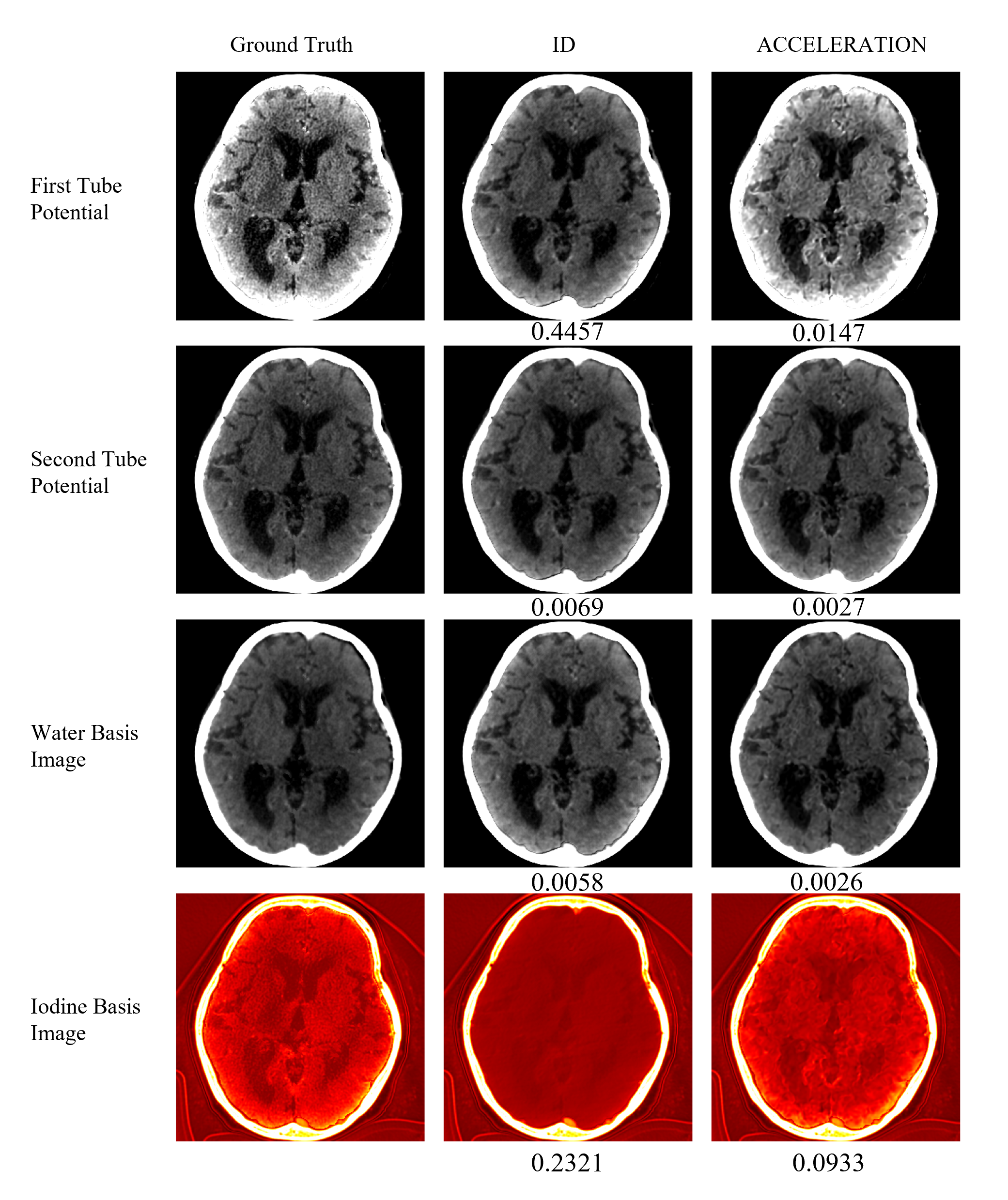}
		\caption{Material basis images. First to third column: ground truth, FBP without extrapolation and ACCELERATION. First to forth rows: CT at low tube potential with display windows [0,120] HU, CT at high tube potential with display windows [0,120] HU, water basis images with display windows [1000,1100] $g/cm^3$ and iodine basis images with display windows [0,15] $g/cm^3$. The numbers below the image represent the NRMSE between the image and ground truth.}
		\label{figure7}
	\end{figure}
	
	In this section, we perform material decomposition based on CT images at two tube potentials that are consistent in time, thereby obtaining relatively accurate material basis images. Fig.~\ref{figure7} shows a comparison experiment of image-domain material decomposition (ID)\footnote{To the best of the authors’ knowledge, this paper presents the high temporal resolution CT imaging based temporal extrapolation algorithm to firstly address the inconsistency in iodine concentration in sequentially-scanning DECT. Therefore, we only compares the proposed technique with direct image-domain material decomposition. In this context, the reconstruction accuracy before material decomposition has a relatively small impact on the accuracy loss caused by inconsistent iodine concentration. Specifically, Filtered Back Projection (FBP) is used as the reconstruction algorithm for comparative experiments.} and material decomposition after extrapolation using ACCELERATION. It can be clearly seen that after the correction of ACCELERATION, the accuracy of material basis images after material decomposition are significantly improved.

	\section{Discussions and Conclusion}
	
	In this work, it has been demonstrated that using ACCELERATION, the technical challenge induced by temporal inconsistency of sequentially-scanned data sets can be mitigated. Results shown in this work demonstrated the improvement of iodine quantification accuracy using ACCELERATION. In future, we plan to incorporate a broader range of ablation studies to investigate the performance dependence of ACCELLERATION. Additionally, we aim to demonstrate the technical feasibility of ACCELLERATION using experimental animal subject studies. It is important to note that the ground truth of iodine concentration of experimental animal subjects who underwent dual-energy contrast-enhanced imaging is very difficult to obtain. More efforts are needed to establish proper references of iodine concentration to consolidate experimental validation using animal subjects. 
	
	%Consequently, this study confines itself to the use of numerical simulation studies because the known ground truth in numerical simulation studies facilitates the demonstration of technical feasibility.}
	
	\bibliographystyle{IEEEbib}
	\bibliography{strings}
	
\end{document}